\documentclass[pra,aps,showpacs,groupedaddress,superscriptaddress,twocolumn,toc=flat]{revtex4-1}
\usepackage[colorlinks=true,linkcolor=blue,urlcolor=blue,citecolor=blue]{hyperref}

\usepackage[utf8x]{inputenc}
\usepackage{color}
\usepackage{bbm} 

\usepackage{amsfonts,amsmath,amssymb,stmaryrd}

\usepackage{graphicx}
\usepackage{subfigure}  
\usepackage{bbm} 
\usepackage{hyperref}
\usepackage{epsfig}
\usepackage{mathrsfs}
\usepackage{verbatim}
\usepackage{centernot}
\usepackage{ulem}
\usepackage{xspace}
\usepackage[english, status=draft]{fixme}       
\fxusetheme{color}                              %

\renewcommand{\l}{\left(}
\renewcommand{\r}{\right)}

\renewcommand{\P}{\hat{P}}
\newcommand{\U}{\hat{U}}

\newcommand{\bra}[1]{\langle#1|}
\newcommand{\ket}[1]{|#1\rangle}

\renewcommand{\ij}{{\langle \vec{i}, \vec{j} \rangle}}
\renewcommand{\H}{\hat{\mathcal{H}}}

\newcommand{\hc}{\text{h.c.}}

\newcommand{\eff}{\text{eff}}

\newcommand{\Cs}{\ensuremath{^{133}{\rm Cs}}\xspace}

\usepackage{array}

\usepackage{cancel,ifthen}
\newcommand{\cmnt}[2][NoInPuT]{\ifthenelse{\equal{#1}{NoInPuT}}{}{{\color{red}\sout{#1}}} {\color{blue} #2}}

\usepackage{bm}	
\renewcommand{\vec}[1]{\bm{#1}}

\begin{document}
\normalem	

\title{Multiparticle interactions for ultracold atoms in optical tweezers:\\ Cyclic ring-exchange terms}

\author{Annabelle Bohrdt}
\thanks{These authors contributed equally.}
\affiliation{Department of Physics and Institute for Advanced Study, Technical University of Munich, 85748 Garching, Germany}
\affiliation{Munich Center for Quantum Science and Technology (MCQST), Schellingstr. 4, D-80799 M\"nchen, Germany}

\author{Ahmed Omran}
\thanks{These authors contributed equally.}
\affiliation{Department of Physics, Harvard University, Cambridge, Massachusetts 02138, USA}

\author{Eugene Demler}
\affiliation{Department of Physics, Harvard University, Cambridge, Massachusetts 02138, USA}

\author{Snir Gazit}
\affiliation{Racah Institute of Physics and The Fritz Haber Research Center for Molecular Dynamics, The Hebrew University, Jerusalem 91904, Israel}

\author{Fabian Grusdt}
\email[Corresponding author email:~]{fabian.grusdt@lmu.de}
\affiliation{Department of Physics and Arnold Sommerfeld Center for Theoretical Physics (ASC), Ludwig-Maximilians-Universit\"at M\"unchen, Theresienstr. 37, M\"unchen D-80333, Germany}
\affiliation{Munich Center for Quantum Science and Technology (MCQST), Schellingstr. 4, D-80799 M\"nchen, Germany}
\affiliation{Department of Physics and Institute for Advanced Study, Technical University of Munich, 85748 Garching, Germany}


\date{\today}

\begin{abstract}
Dominant multi-particle interactions can give rise to exotic physical phases with anyonic excitations and phase transitions without local order parameters. In spin systems with a global $SU(N)$ symmetry, cyclic ring-exchange couplings constitute the first higher-order interaction in this class. In this letter we propose a protocol how $SU(N)$ invariant multi-body interactions can be implemented in optical tweezer arrays. We utilize the flexibility to re-arrange the tweezer configuration on time scales short compared to the typical lifetimes, in combination with strong non-local Rydberg interactions. As a specific example we demonstrate how a chiral cyclic ring-exchange Hamiltonian can be implemented in a two-leg ladder geometry. We study its phase diagram using DMRG simulations and identify phases with dominant vector chirality, a ferromagnet, and an emergent spin-$1$ Haldane phase. We also discuss how the proposed protocol can be utilized to implement the strongly frustrated $J-Q$ model, a candidate for hosting a deconfined quantum critical point.
\end{abstract}

\maketitle

\emph{Introduction.--}
Ultracold atoms in optical lattices have become a versatile platform for performing analogue quantum simulations, with widely tunable interactions \cite{Bloch2008} and the ability to control the single-particle band structure \cite{Tarruell2012,Lin2011,Aidelsburger2011,Cheuk2012,Struck2012,Li2016PRL,Wu2016}. Using atoms with permanent electric or magnetic dipole moments \cite{Lahaye2009} or in Rydberg states \cite{Saffman2010} allows to study systems with long-range dipole-dipole or van-der Waals interactions, which can mimic the long-range Coulomb repulsion between electrons in a solid. These ingredients can be combined to study exotic phenomena in strongly correlated many-body systems, related for example to quantum magnetism \cite{Simon2011,Greif2013,Hild2014,Hart2015,Hilker2017,Mazurenko2017,Brown2017} or the fractional quantum Hall effect \cite{Gemelke2010,Tai2017,Clark2019}. Leveraging the capabilities of ultracold atoms, such experiments promise new insights for example to directly measure topological invariants \cite{Atala2012,Grusdt2016TP,Flaschner2016,Li2016,Asteria2019} or image the quantum mechanical wavefunction with single-site resolution \cite{Gericke2008,Bakr2009,Sherson2010,Parsons2016,Boll2016,Cheuk2016}.

\begin{figure}[b!]
\centering
\epsfig{file=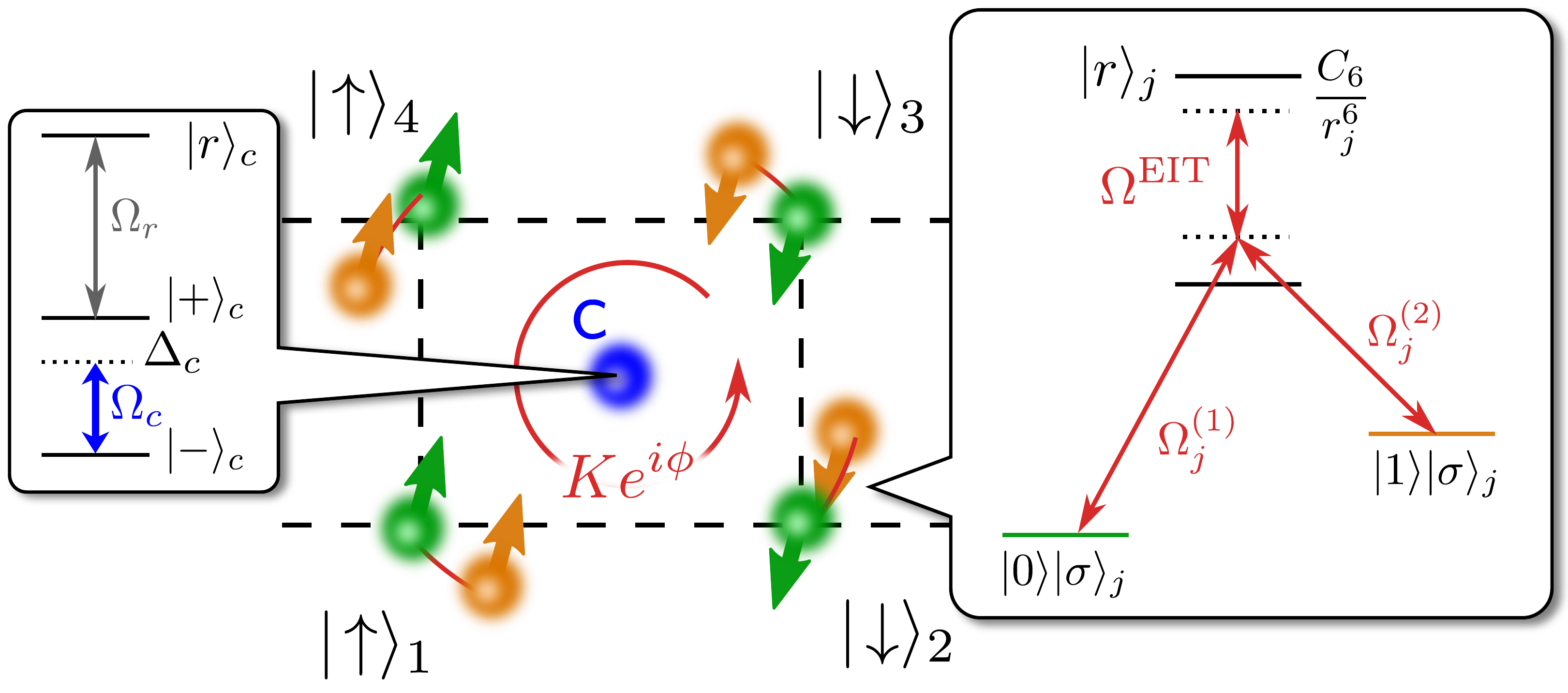, width=0.49\textwidth}
\caption{Proposed setup: SU$(2)$-invariant chiral cyclic ring-exchange interactions can be realized by combining state-dependent lattices generated by optical tweezer arrays and strong Rydberg interactions with a central Rydberg-dressed control qubit (C). The auxiliary states $\ket{\tau=1}\ket{\sigma}$ with $\sigma = \uparrow, \downarrow$ (orange) of the atoms on the sites of the plaquette are subject to a state-dependent tweezer potential which allows to permute them coherently around the center. Our protocol makes use of stroboscopic $\pi$ pulses between the physical states $\tau=0$ (green) and the auxiliary states $\tau=1$, which only take place collectively on all sites and conditioned on the absence of a Rydberg excitation in the control atom.} 
\label{figSetup}
\end{figure}

In this letter, we go beyond the two-body interactions realized so far and propose a general protocol to implement highly symmetric multiparticle interactions with ultracold atoms in optical tweezer arrays. Multiparticle interactions can lead to exotic ground states with intrinsic topological order \cite{Read1999,Kitaev2006a}, with applications for quantum computation \cite{Kitaev2003,Nayak2008}, and they are an important ingredient for realizing lattice gauge theories \cite{Buchler2005,Tagliacozzo2012,Wiese2013,Zohar2016} central to the quantum simulation of high-energy phenomena or deconfined quantum criticality \cite{Senthil2004,Sandvik2007}. If these higher-order couplings possess additional symmetries, e.g. SU$(N)$ invariance in spin systems, models with strong frustration can be realized whose ground states are strongly correlated quantum liquids. 

In condensed matter systems multi-spin interactions of this type emerge from higher-order virtual processes \cite{Dirac1929}, leading to corrections to the pairwise Heisenberg couplings of SU$(2)$ spins in a half-filled Hubbard model. These cyclic ring-exchange terms play a role in frustrated quantum magnets like solid $~^{3}{\rm He}$ \cite{Roger1983} and possibly also for the phase diagram of high-$T_{\rm c}$ cuprate superconductors \cite{Roger1989,Roger2005}. In this letter we demonstrate how such multi-spin interactions can be realized and independently tuned in ultracold atom systems without resorting to high-order virtual processes. 

A promising route to implementing multiparticle processes is to use strong interactions between atoms in different Rydberg states representing spin degrees of freedom. This allows to build a versatile quantum simulator which can be used to realize ring-exchange interactions in spin systems by representing them as sums of products of Pauli matrices \cite{Weimer2010}, or to implement local constraints giving rise to emergent dynamical gauge fields \cite{Glaetzle2014,Surace2019}.

Here we follow a similar strategy but propose to combine strong Rydberg interactions with the capabilities to quickly change the spatial configuration of atoms trapped in optical tweezer arrays \cite{Barredo2016,Endres2016,Kim2016}. We consider general lattice models with one $N$-component particle per lattice site (fermionic or bosonic) and show, as an explicit example, how a general class of SU$(N)$-invariant chiral cyclic ring-exchange (CCR) interactions can be realized. They are described by a Hamiltonian ($\hbar=1$)
\begin{equation}
\H_{\rm CCR}(\phi) =  K \sum_p ( e^{i \phi} \P_p + e^{ - i \phi}  \P^\dagger_p).
\label{eqCCRdef}
\end{equation}
where the sum is over all plaquettes $p$ of the underlying lattice, the operator $\hat{P}_p^\dagger$ ($\hat{P}_p$) cyclically permutes the spin configuration on plaquette $p$ in clockwise (counterclockwise) direction and $\phi$ is a tunable complex phase. A generalization to finite hole doping, with zero or one particle per lattice site, is straightforward.

Non-chiral cyclic ring-exchange interactions, realized by Eq.~\eqref{eqCCRdef} for $\phi=0$, are believed to play a role in the high-$T_{\rm c}$ cuprate compounds. These materials can be described by the 2D Fermi-Hubbard model on a square lattice, with weak couplings between multiple layers in $z$-direction \cite{Emery1987}. For the relevant on-site interactions $U$, which dominate over the nearest-neighbor tunneling $t \ll U$, this model can be simplified by an expansion in powers of $t/U$. To lowest order, one obtains a $t-J$ model \cite{Auerbach1998} with nearest-neighbor spin-exchange interactions of strength $J=4 t^2/U$. Next to leading order, cyclic ring-exchange terms on the plaquettes of the square lattice contribute with strength $K=20 t^4/U^3$. By comparison of first principle calculations and measurements in the high-temperature regime it was shown that $K \approx 0.13 \times J$ in ${\rm La}_2 {\rm Cu} {\rm O}_4$ \cite{Toader2005} but its effect on the phase diagram remains debated. In ultracold atoms, similar higher-order processes have been used to realize non-chiral cyclic ring-exchange couplings \cite{Paredes2008,Dai2017}.

We start by explaining the general scheme using the example of CCR interactions. Our method is more versatile however, and we discuss how it can be adapted to implement the $J-Q$ model which has been proposed as a candidate system realizing deconfined quantum criticality \cite{Senthil2004,Sandvik2007}. We also analyze the phase diagram of the CCR Hamiltonian \eqref{eqCCRdef} in a ladder geometry, with exactly one SU$(2)$ spin per lattice site. We show that the phase diagram contains a gapped Haldane phase with topologically protected edge states \cite{Haldane1983b,Kennedy1990,Kennedy1992} at intermediate values of $\pi/4 \lesssim \phi \lesssim 3\pi/4$, a ferromagnetic phase for $\phi \gtrsim 3\pi/4$ and a dominant vector chirality for $\phi \lesssim \pi/4$.

\emph{Implementation.--}
For simplicity we consider only a single plaquette and for concreteness we restrict ourselves to $N_p=4$ sites, see Fig.~\ref{figSetup}. Generalizations of our scheme to more than one plaquette with any number $N_p$ of sites are straightforward, however. 

Each of the four sites, labeled $j=1,...,4$, consists of a static optical tweezer trapping a single atom, where recently demonstrated rearrangement methods~\cite{Barredo2016, Endres2016, Kim2016} allow for populating each site with high fidelity. We assume that the atoms remain in the vibrational ground states of the microtraps throughout the sequence. Every atom has two internal states $\sigma = \uparrow, \downarrow$ which we use to implement an effective spin-$1/2$ system. As a specific configuration we suggest to use \Cs atoms and utilize their $F=3$, $m_F=2,3$ hyperfine states to represent the two spins. Optical pumping with site-resolved addressing can then be employed to prepare arbitrary initial spin patterns~\cite{Dai2017} and study their dynamics under Eq.~\eqref{eqCCRdef}.

The key ingredient for our proposed implementation of CCR interactions is to realize collective permutations of the entire spin configuration in the plaquette. This can be achieved by physically rotating the tweezer array around the center of the plaquette while ensuring that the motional and spin states of the atoms are preserved and coherence is not lost. The effect of clockwise rotations of the microtraps on the spin states is described by the operator $\P$,
\begin{equation}
\P  \ket{\sigma_1, \sigma_2, \sigma_3, \sigma_4} = \ket{\sigma_4, \sigma_1, \sigma_2, \sigma_3}. 
\label{eqDefPp}
\end{equation}
Optimized trajectories can be chosen to cancel heating effects from the motion~\cite{Murphy2009}. These require a timescale set by the quantum speed limit that scales as the inverse energy gap of each traps $t_{\rm rot}\sim 1/\Delta \varepsilon$.  For deep trapping potentials where $\Delta \varepsilon\approx 150\,$kHz, rotation times of $t_{\rm rot}<10\,\mu$s are achievable.

In contrast to Eq.~\eqref{eqDefPp}, the effective Hamiltonian leads to a superposition of permuted and non-permuted states in every infinitesimal time step $\Delta t$, as can be seen from a Taylor expansion: $e^{- i \H_{\rm CCR} \Delta t} = 1 - i \H_{\rm CCR} \Delta t$. To create such superposition states in our time evolution, we assume that every atom has a second internal degree of freedom labeled by $\tau = 0,1$. Concretely we propose to realize the new states $\ket{\tau=1} \ket{\sigma}$ in \Cs atoms by $F=4$, $m_F=3,4$ hyperfine levels, where $m_F=3$ ($m_F=4$) corresponds to $\sigma=\downarrow$ ($\sigma=\uparrow$). These additional levels will be used as auxiliary states, whereas the states $\ket{\tau=0} \ket{\sigma}$ introduced before -- implemented as $F=3$, $m_F=2,3$ levels in \Cs -- realize the physical spin states. 

One part of our protocol consists of a permutation of the spins $\sigma$, but only in the manifold of auxiliary states. This step requires a total time $t_{\rm rot}$ and can be described by the unitary transformations
\begin{equation}
\U_{+} = \prod_j \ket{1}_j \bra{1} \otimes \P ~ +~  \prod_j \ket{0}_j \bra{0} \otimes \hat{\mathbf{1}}_\sigma, \quad \U_{-} =  \U_{+}^\dagger \label{eqDefUpls}
\end{equation}
To implement this evolution, two sets of optical tweezer arrays can be used, of which only one is rotating. We suggest to realize it by the near-magic wavelength $\lambda_{\rm magic}\approx 871.6\,$nm in \Cs which strongly confines atoms in the state $\tau = 1$ but almost does not affect atoms in $\tau=0$. By applying $\U_{\pm}$ to superposition states with either all atoms in $\tau=1$ or all atoms in $\tau=0$, one can realize the desired superpositions of permuted and non-permuted spin configurations. Such states can be realized by collective $\pi$-pulses conditioned upon a control qubit trapped in the center of the plaquette~\cite{Muller2009}, as described next.

If the control atom is in the state $\ket{+}_c$ it is transferred to a Rydberg state $\ket{r}_c$ with a resonant $\pi$-pulse and Rabi frequency $\Omega_r$, see Fig.~\ref{figSetup}. If the control atom is in state $\ket{-}_c$, the laser $\Omega_r$ is off-resonant and no Rydberg excitation is created. Next a Raman transition by lasers $\Omega^{(1)}_j$, $\Omega^{(2)}_j$ through an intermediate Rydberg state $\ket{r}_j$ is used to implement a $\pi$-pulse transferring the physical states $\ket{0}_j$ to $\ket{1}_j$, without changing their spin state $\ket{\sigma}_j$. In the presence of a coupling field $\Omega^{\rm EIT}$ that establishes two-photon resonance to the Rydberg state with each Raman laser, electromagnetically induced transparency (EIT) \cite{Fleischhauer2005} suppresses the transition $\ket{0}_j \leftrightarrow \ket{1}_j$. However, the EIT condition is lifted by the Rydberg blockade mechanism if the control atom is in the Rydberg state $\ket{r}_c$~\cite{Muller2009}, enabling the transfer. After the transfer is complete, another $\pi$-pulse by $\Omega_r$ is applied to the control atom. This ensures that the control atom remains trapped during the protocol, even if the Rydberg excited state is not subject to a trapping potential. In summary, this part is described by the unitary transformation
\begin{multline}
\U_{\rm sw} = \ket{+}_c \bra{+} \otimes \biggl( \prod_j  \ket{1}_j \bra{0} + \hc \biggr) \otimes \hat{\mathbf{1}}_\sigma \\
+  \ket{-}_c \bra{-} \otimes \hat{\mathbf{1}}_\tau \otimes \hat{\mathbf{1}}_\sigma.
\label{eqUsw}
\end{multline}
The total time required to implement this switch (sw) is denoted by $t_{\rm sw}$.

Finally, we need to introduce quantum dynamics between the states of the control atom. This can be realized by a dressing laser $\Omega_c$ driving transitions between $\ket{\pm}_c$, at a detuning $\Delta_c$. These dynamics take place over a period of time $t_{c}$ and are described by the unitary evolution $\U_c = e^{- i \H_c t_c}$ with $\H_c = \Delta_c \ket{+}_c \bra{+} + \Omega_c \bigl( \ket{+}_c \bra{-} + \hc \bigr)$. During the remaining steps of the protocol, Eqs.~\eqref{eqDefUpls} - \eqref{eqUsw}, we assume that $\Omega_c = 0$ is off and the control atom picks up a phase $\pm \varphi_c$ if it is in state $\ket{+}_c$. This phase can be adjusted by the detuning $\Delta_c$ and the duration $t_{\rm rx} = 2 t_{\rm sw} + t_{\rm rot}$, during which the time evolution of the control is $\U_{\pm \varphi_c} = \ket{+}_c \bra{+} e^{ \mp i \varphi_c} + \ket{-}_c \bra{-}$.

\begin{figure}[t!]
\centering
\epsfig{file=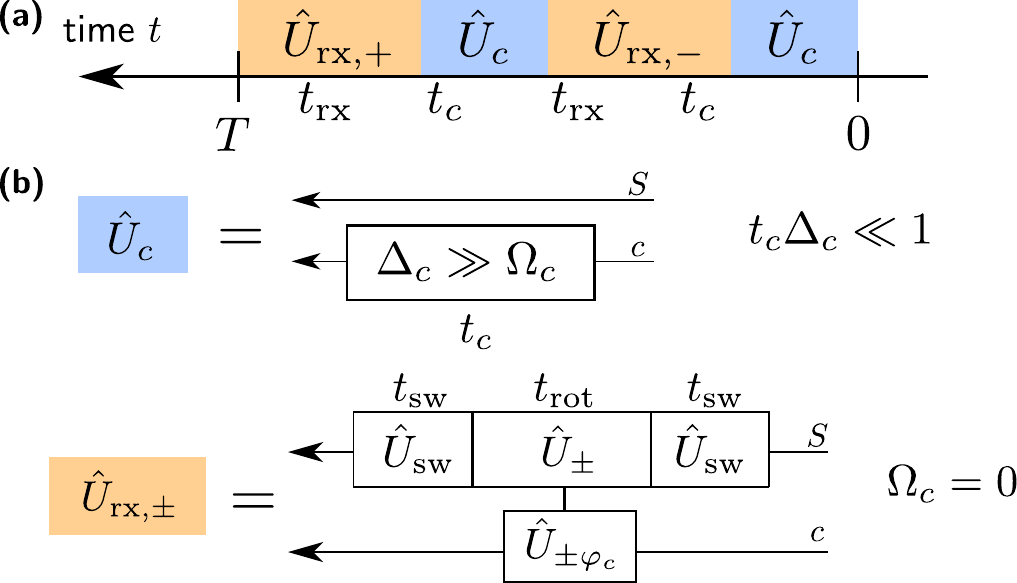, width=0.45\textwidth}
\caption{Proposed protocol: The sequence in (a) is repeated periodically with period $T = 2 (t_c + t_{\rm rx})$. When $t_c \ll 2 \pi/\Delta_c$, $1/\Omega_c$ it implements a trotterized time evolution of the effective Hamiltonian Eq.~\eqref{eqDefHeffScheme}, which realizes CCR couplings when $\Delta_c \gg \Omega_c$. The individual time steps are illustrated in (b).} 
\label{figProtocol}
\end{figure}

The complete protocol is summarized in Fig.~\ref{figProtocol}. It consists of a periodic repetition of the individual steps described above. At the discrete time steps $n T$, where $T = 2 (t_c + t_{\rm rx})$, the unitary evolution is described by an effective Hamiltonian $\H_\eff$:
\begin{equation}
e^{- i n T \H_\eff }  = (\U_T)^n  = \l \U_{{\rm rx},+} \U_c \U_{{\rm rx},-} \U_c \r^n,
\end{equation}
where we defined $ \U_{{\rm rx},\pm} = \U_{\rm sw} \l \U_{\pm \varphi_c} \otimes \U_{\pm} \r \U_{\rm sw}$. As will be shown below, $\H_\eff$ realizes CCR interactions with a tunable phase $\phi=-\varphi_c$ and amplitude
\begin{equation}
K = - \frac{1}{2T}  (t_c \Delta_c) \l \frac{\Omega_c}{\Delta_c} \r^2
\end{equation}
provided that 
\begin{equation}
t_c \ll 2 \pi/\Delta_c, \qquad \Omega_c \ll \Delta_c.
\label{eqCond}
\end{equation}

Now we estimate the strength $|K|$ of the CCR interactions that can be achieved with the proposed setup. To satisfy Eq.~\eqref{eqCond} we assume $\Omega_c = 0.2 \Delta_c$ and $t_c \Delta_c = 0.4$. For a rotation time $t_{\rm rot} = 10 \mu {\rm s}$ and assuming $t_{\rm sw}, t_{c} \ll t_{\rm rot}$ a reasonable strength of $K / \hbar =  50 {\rm Hz} \times 2 \pi$ can be achieved. This requires $\Omega_c / 2 \pi \gg 1.3 {\rm kHz}$, which can be easily realized; the condition $t_{\rm sw} \ll 10 \mu {\rm s}$ can also be met, as the Rydberg $\pi$-pulses on the control atom can be executed in $\sim100$ ns each and the Raman transfer between the states $\ket{0}_j$ and $\ket{1}_j$ can be driven with coupling strengths above $1$ MHz.

\emph{Effective Hamiltonian.--}
Next we show that our protocol realizes the Hamiltonian in Eq.~\eqref{eqCCRdef}. When $2 \pi t_c \ll 1/\Delta_c, 1/\Omega_c$, we can write $\U_c = 1 - i \H_c t_c$ and calculate $\exp[ -i \H_\eff T]$ to leading order in $t_c$.  Eqs.~\eqref{eqDefUpls} - \eqref{eqUsw} yield
\begin{multline}
\H_{\eff} = \frac{t_c}{T} \biggl\{ 2 \Delta_c \ket{+}_c \bra{+} + \\
+  \Omega_c \left[ \ket{-}_c\bra{+} \l 1 + e^{i \varphi_c} \P^\dagger \r + \hc \right] \biggr\}.
\label{eqDefHeffScheme}
\end{multline}

When $\Omega_c \ll \Delta_c$ we can eliminate the state $\ket{+}_c$ which is only virtually excited. This further simplifies the effective Hamiltonian and we obtain
\begin{equation}
\H_\eff = K \l  2 + e^{- i \varphi_c} \P + e^{ i \varphi_c} \P^\dagger  \r.
\label{eqHeffCCRderivation}
\end{equation}
Up to the energy shift $2K$ this realizes CCR interactions in an isolated plaquette. The result can be extended to multiple plaquettes by implementing the trotterized time step $\U_T$ interchangeably on inequivalent plaquettes.

\begin{figure}[t!]
\centering
\epsfig{file=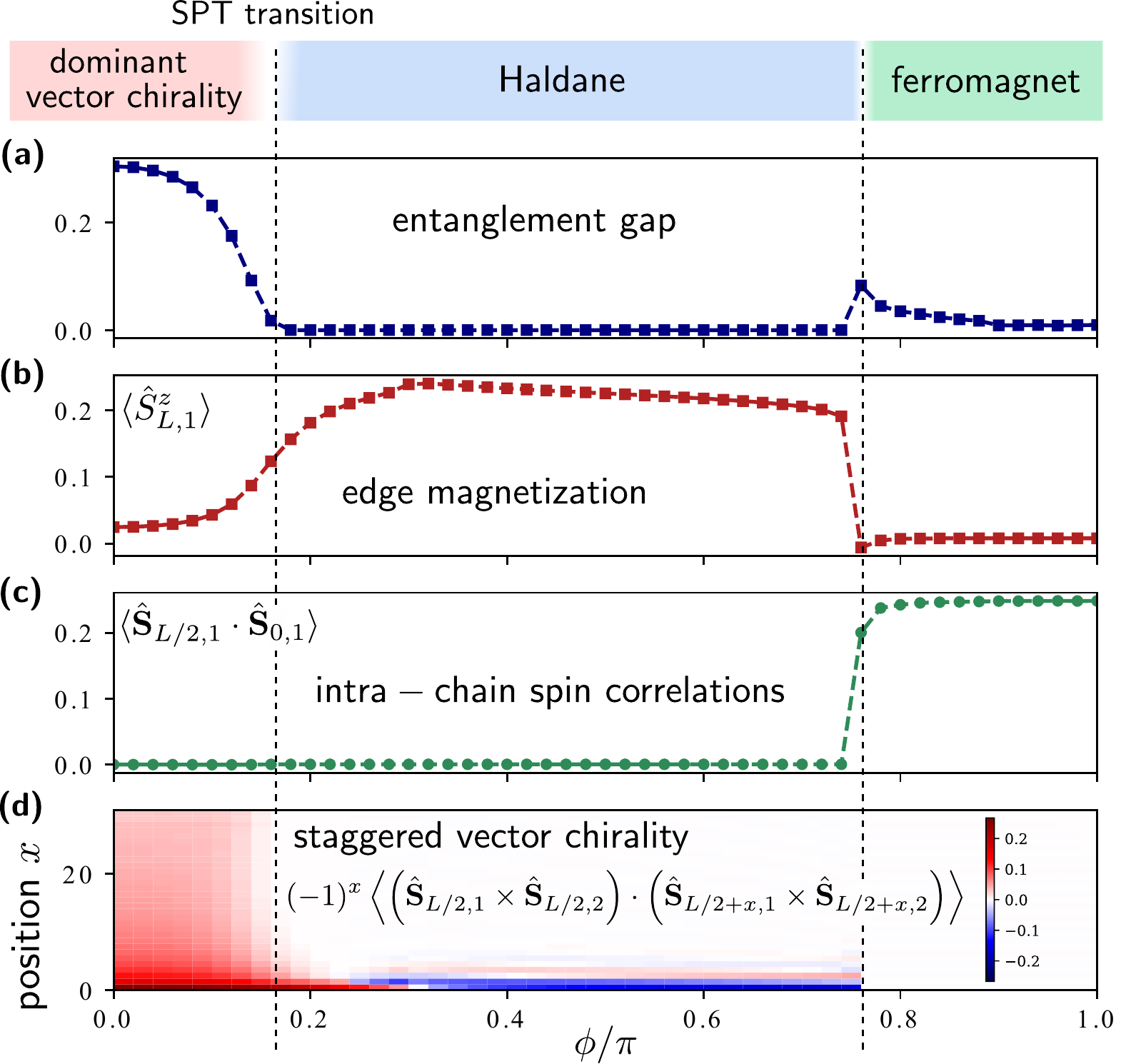, width=0.49\textwidth}
\caption{Phase diagram of the CCR Hamiltonian on a ladder, obtained from DMRG in a system with 64 sites: different observables are evaluated in the ground state of the Hamiltonian \eqref{eqCCRdef} to characterize the phases. Upon varying $\phi$, three different phases can be identified: A topological Haldane phase featuring a vanishing gap in the entanglement spectrum (a) and edge states with a non-zero local magnetization for $S^z_{\rm tot}=1$ (b); A symmetry-broken phase around $\phi = \pi$ with long-range ferromagnetic correlations (c); And a symmetric phase for small $\phi$, where the staggered vector chirality remains non-vanishing over long distances (d).} 
\label{figPhaseDiag}
\end{figure}

\emph{Two-leg ladder with CCR.--}
Now we discuss the physics of the CCR Hamiltonian in a two-leg ladder. We vary the phase $\phi$ in the Hamiltonian \eqref{eqCCRdef} with $K=1$ and calculate the ground state phase diagram using the density-matrix renormalization group (DMRG). 
For $\phi = \pi$, the ground state is characterized by ferromagnetic correlations, see Fig.~\ref{figPhaseDiag} (c). It can be readily seen that the variational energy $\langle \H_{\rm CCR}(\pi) \rangle$ is minimized for ferromagnetic configurations. In the sector $S^z_{\rm tot} = 0$ used in our DMRG simulations in Fig.~\ref{figPhaseDiag} (c), we find phase separation with two ferromagnetic domains of opposite magnetization.

At intermediate $\phi$ we find an emergent Haldane phase, with two-fold degenerate states in the entanglement spectrum, see Fig.~\ref{figPhaseDiag} (a). For a finite $S^z_{\rm tot} = 1$ the expectation value $\langle \hat{S}^z_{L,1}\rangle$ at the edge is non-zero, see Fig.~\ref{figPhaseDiag} (b). The spin gap $\Delta E_S = E_{0,S=1} - E_{0,S=0}$, defined as the difference between the ground state energy with and without finite total magnetization, is zero in this phase, since the additional spin can be placed in the spin-$1/2$ topological edge states of the system without increasing the total energy. We corroborate this picture further by considering the $K-K'$ model with alternating strengths $K$, $K'$ of the CCR interactions on adjacent plaquettes. In the supplement (SM) we provide an explicitly derivation of a spin-$1$ model with a gapped Haldane ground state \cite{Haldane1983a,Kennedy1990} for $\phi=\pi/2$ and $K' \ll K$.

For small values of $\phi$, the ground state of the CCR Hamiltonian is a dominant vector chirality phase, as discussed in Ref.~\cite{Laeuchli2003}. This phase is characterized by correlations of the form $\hat{\vec{S}}_{x,y} \times \hat{\vec{S}}_{x',y'}$ in a staggered arrangement around each plaquette. We find that the staggered correlation between different rungs, measured from the center $L/2$ of the chain,
\begin{equation}
(-1)^x \left\langle \left( \hat{\vec{S}}_{L/2,1} \times \hat{\vec{S}}_{L/2,2}\right) \cdot  \left( \hat{\vec{S}}_{L/2+x,1} \times \hat{\vec{S}}_{L/2+x,2}\right) \right\rangle,
\end{equation}
decays slowly as a function of the distance $x$ and retains significant non-zero values over the considered system sizes, see Fig.~\ref{figPhaseDiag} (d). The transition between the dominant vector chirality and Haldane phases is a symmetry-protected topological (SPT) phase transition. 

Using the global $SU(2)$ symmetry, the staggered vector chirality becomes $6 \Bigl\langle  \hat{S}_{L/2,1}^x  \hat{S}_{L/2,2}^y   \Bigl(  \hat{S}_{L/2+x,1}^x \hat{S}^y_{L/2+x,2}$ $ - \hat{S}_{L/2+x,1}^y \hat{S}^x_{L/2+x,2}  \Bigr) \Bigr\rangle (-1)^x $. Measuring it requires access to two four-point functions of the form $\langle \hat{S}^\mu_{\vec{i}} \hat{S}^\nu_{\vec{j}} \hat{S}^\lambda_{\vec{k}} \hat{S}^\rho_{\vec{l}} \rangle$ which can be detected by making use of local addressing techniques, see e.g. \cite{Weitenberg2011}. To detect the Haldane phase experimentally, we propose to study weakly magnetized systems and image the topological edge states. Alternatively, one could work in the plaquette basis (see SM) and measure the Haldane string order parameter. An interesting future extension would be to use machine learning techniques to retrieve non-local order parameters from a series of quantum projective measurements. 

\emph{Summary and Outlook.--}
In summary, we propose a general method for implementing multi-body interactions in ultracold atom experiments using optical tweezer arrays. The approach is particularly useful in the presence of additional, e.g. global $SU(N)$ spin, symmetries. Specifically, we consider a four-body cyclic ring exchange term, which can be realized with a combination of multi-qubit gates based on Rydberg states and movable optical tweezers. We numerically study the ground state of the cyclic ring exchange Hamiltonian and find different dominant correlation functions as the complex phase of the ring exchange term is varied. 

Our work paves the way for future studies of the interplay between ring-exchange and pair-exchange terms, as discussed in Ref.~\cite{Metavitsiadis2017} for the non-chiral case $\phi=0$. In the experimental realization proposed here, it is conceptually straightforward to introduce holes into the system, leading to a finite doping. The interplay between spin and charge degrees of freedom could be further studied by adding direct tunneling terms, which lead to rich Hamiltonians in the spirit of $t-J$ like models. The physics of this type of model is completely unknown and provides an exciting prospect for future theoretical and experimental research. 
The proposed protocol is versatile enough to implement larger classes of models with multi-spin interactions, such as the $J-Q$ model \cite{Sandvik2007}. In two dimensions, this model features a phase transition between an antiferromagnet and a valence-bond solid, which has been proposed as a candidate for a deconfined quantum critical point \cite{Sandvik2007}. 
Moreover, the experimental protocol can be varied to study different types of problems, such as discrete time-evolutions of complex models or impurity models, which can be realized by an inclusion of the control qubits into the models. 

\emph{Acknowledgements.--}
The authors would like to thank T. Calarco, M. Endres, M. Greiner, A. Kaufman, M. Knap and M. Lukin for fruitful discussions. A.B. and F.G. acknowledge support from the Technical University of Munich - Institute for Advanced Study, funded by the German Excellence Initiative and the European Union FP7 under grant agreement 291763, the Deutsche Forschungsgemeinschaft (DFG, German Research Foundation) under Germany's Excellence Strategy--EXC-2111--390814868, DFG grant No. KN1254/1-1, DFG TRR80 (Project F8). A.B. acknowledges support by the Studienstiftung des deutschen Volkes. A.O. acknowledges support by a research fellowship from the German Research Foundation (DFG). E.D. acknowledges support by Harvard-MIT CUA, AFOSR-MURI Quantum Phases of Matter (grant FA9550-14-1-0035), AFOSR-MURI: Photonic Quantum Matter (award FA95501610323), DARPA DRINQS program (award D18AC00014). S.G. acknowledges support from the Israel Science Foundation, Grant No. 1686/18. F.G. acknowledges support by the Gordon and Betty Moore foundation through the EPiQS program.


\begin{thebibliography}{}

\end{thebibliography}


\begin{thebibliography}{10}

\bibitem{Bloch2008}
Immanuel Bloch, Jean Dalibard, and Wilhelm Zwerger.
\newblock Many-body physics with ultracold gases.
\newblock {\em Reviews of Modern Physics}, 80(3):885--964, 2008.

\bibitem{Tarruell2012}
Leticia Tarruell, Daniel Greif, Thomas Uehlinger, Gregor Jotzu, and Tilman
  Esslinger.
\newblock Creating, moving and merging dirac points with a fermi gas in a
  tunable honeycomb lattice.
\newblock {\em Nature}, 483(7389):302--U91, March 2012.

\bibitem{Lin2011}
Y.-.~J. Lin, K.~Jimenez-Garcia, and I.~B. Spielman.
\newblock Spin-orbit-coupled bose-einstein condensates.
\newblock {\em Nature}, 471(7336):83--U99, March 2011.

\bibitem{Aidelsburger2011}
M.~Aidelsburger, M.~Atala, S.~Nascimbene, S.~Trotzky, Y.-.~A. Chen, and
  I.~Bloch.
\newblock Experimental realization of strong effective magnetic fields in an
  optical lattice.
\newblock {\em Physical Review Letters}, 107(25):255301, December 2011.

\bibitem{Cheuk2012}
Lawrence~W. Cheuk, Ariel~T. Sommer, Zoran Hadzibabic, Tarik Yefsah, Waseem~S.
  Bakr, and Martin~W. Zwierlein.
\newblock Spin-injection spectroscopy of a spin-orbit coupled fermi gas.
\newblock {\em PRL}, 109:095302, 2012.

\bibitem{Struck2012}
J.~Struck, C.~\"Olschl\"ager, M.~Weinberg, P.~Hauke, J.~Simonet, A.~Eckardt,
  M.~Lewenstein, K.~Sengstock, and P.~Windpassinger.
\newblock Tunable gauge potential for neutral and spinless particles in driven
  optical lattices.
\newblock {\em Phys. Rev. Lett.}, 108:225304, May 2012.

\bibitem{Li2016PRL}
Junru Li, Wujie Huang, Boris Shteynas, Sean Burchesky, Furkan~Cagri Top, Edward
  Su, Jeongwon Lee, Alan~O. Jamison, and Wolfgang Ketterle.
\newblock Spin-orbit coupling and spin textures in optical superlattices.
\newblock {\em Phys. Rev. Lett.}, 117:185301, Oct 2016.

\bibitem{Wu2016}
Zhan Wu, Long Zhang, Wei Sun, Xiao-Tian Xu, Bao-Zong Wang, Si-Cong Ji, Youjin
  Deng, Shuai Chen, Xiong-Jun Liu, and Jian-Wei Pan.
\newblock Realization of two-dimensional spin-orbit coupling for bose-einstein
  condensates.
\newblock {\em Science}, 354(6308):83--88, 2016.

\bibitem{Lahaye2009}
T.~Lahaye, C.~Menotti, Santos L., M.~Lewenstein, and T.~Pfau.
\newblock The physics of dipolar bosonic quantum gases.
\newblock {\em Reports On Progress In Physics}, 72(12):126401, 2009.

\bibitem{Saffman2010}
M.~Saffman, T.~G. Walker, and K.~Molmer.
\newblock Quantum information with rydberg atoms.
\newblock {\em Reviews of Modern Physics}, 82(3):2313--2363, 2010.

\bibitem{Simon2011}
Jonathan Simon, Waseem~S. Bakr, Ruichao Ma, M.~Eric Tai, Philipp~M. Preiss, and
  Markus Greiner.
\newblock Quantum simulation of antiferromagnetic spin chains in an optical
  lattice.
\newblock {\em Nature}, 472(7343):307--312, April 2011.

\bibitem{Greif2013}
Daniel Greif, Thomas Uehlinger, Gregor Jotzu, Leticia Tarruell, and Tilman
  Esslinger.
\newblock Short-range quantum magnetism of ultracold fermions in an optical
  lattice.
\newblock {\em Science}, 340(6138):1307--1310, 2013.

\bibitem{Hild2014}
Sebastian Hild, Takeshi Fukuhara, Peter Schau\ss{}, Johannes Zeiher, Michael
  Knap, Eugene Demler, Immanuel Bloch, and Christian Gross.
\newblock Far-from-equilibrium spin transport in heisenberg quantum magnets.
\newblock {\em Phys. Rev. Lett.}, 113:147205, Oct 2014.

\bibitem{Hart2015}
Russell~A. Hart, Pedro~M. Duarte, Tsung-Lin Yang, Xinxing Liu, Thereza Paiva,
  Ehsan Khatami, Richard~T. Scalettar, Nandini Trivedi, David~A. Huse, and
  Randall~G. Hulet.
\newblock Observation of antiferromagnetic correlations in the hubbard model
  with ultracold atoms.
\newblock {\em Nature}, 519(7542):211--214, March 2015.

\bibitem{Hilker2017}
Timon~A. Hilker, Guillaume Salomon, Fabian Grusdt, Ahmed Omran, Martin Boll,
  Eugene Demler, Immanuel Bloch, and Christian Gross.
\newblock Revealing hidden antiferromagnetic correlations in doped hubbard
  chains via string correlators.
\newblock {\em Science}, 357(6350):484--487, 2017.

\bibitem{Mazurenko2017}
Anton Mazurenko, Christie~S. Chiu, Geoffrey Ji, Maxwell~F. Parsons, Marton
  Kanasz-Nagy, Richard Schmidt, Fabian Grusdt, Eugene Demler, Daniel Greif, and
  Markus Greiner.
\newblock A cold-atom fermi-hubbard antiferromagnet.
\newblock {\em Nature}, 545(7655):462--466, May 2017.

\bibitem{Brown2017}
Peter~T. Brown, Debayan Mitra, Elmer Guardado-Sanchez, Peter Schau\ss,
  Stanimir~S. Kondov, Ehsan Khatami, Thereza Paiva, Nandini Trivedi, David~A.
  Huse, and Waseem~S. Bakr.
\newblock Spin-imbalance in a 2d fermi-hubbard system.
\newblock {\em Science}, 357(6358):1385--, September 2017.

\bibitem{Gemelke2010}
Nathan Gemelke, Edina Sarajlic, and Steven Chu.
\newblock Rotating few-body atomic systems in the fractional quantum hall
  regime.
\newblock {\em arXiv:1007.2677v1}, 2010.

\bibitem{Tai2017}
M.~Eric Tai, Alexander Lukin, Matthew Rispoli, Robert Schittko, Tim Menke, Dan
  Borgnia, Philipp~M. Preiss, Fabian Grusdt, Adam~M. Kaufman, and Markus
  Greiner.
\newblock Microscopy of the interacting harper-hofstadter model in the two-body
  limit.
\newblock {\em Nature}, 546(7659):519--523, June 2017.

\bibitem{Clark2019}
Logan~W. Clark, Nathan Schine, Claire Baum, Ningyuan Jia, and Jonathan Simon.
\newblock Observation of laughlin states made of light.
\newblock {\em arXiv:1907.05872}, 2019.

\bibitem{Atala2012}
Marcos Atala, Monika Aidelsburger, Julio~T. Barreiro, Dmitry Abanin, Takuya
  Kitagawa, Eugene Demler, and Immanuel Bloch.
\newblock Direct measurement of the zak phase in topological bloch bands.
\newblock {\em Nature Physics}, 9:795, 2013.

\bibitem{Grusdt2016TP}
F.~Grusdt, N.~Y. Yao, D.~Abanin, M.~Fleischhauer, and E.~Demler.
\newblock Interferometric measurements of many-body topological invariants
  using mobile impurities.
\newblock {\em Nat Commun}, 7:11994, June 2016.

\bibitem{Flaschner2016}
N.~Fl\"aschner, B.~S. Rem, M.~Tarnowski, D.~Vogel, D.-S. L\"uhmann,
  K.~Sengstock, and C.~Weitenberg.
\newblock Experimental reconstruction of the berry curvature in a floquet bloch
  band.
\newblock {\em Science}, 352(6289):1091--1094, 2016.

\bibitem{Li2016}
Tracy Li, Lucia Duca, Martin Reitter, Fabian Grusdt, Eugene Demler, Manuel
  Endres, Monika Schleier-Smith, Immanuel Bloch, and Ulrich Schneider.
\newblock Bloch state tomography using wilson lines.
\newblock {\em Science}, 352(6289):1094--1097, 2016.

\bibitem{Asteria2019}
Luca Asteria, Duc~Thanh Tran, Tomoki Ozawa, Matthias Tarnowski, Benno~S. Rem,
  Nick Fl\"aschner, Klaus Sengstock, Nathan Goldman, and Christof Weitenberg.
\newblock Measuring quantized circular dichroism in ultracold topological
  matter.
\newblock {\em Nature Physics}, 15(5):449--454, 2019.

\bibitem{Gericke2008}
Tatjana Gericke, Peter Wuertz, Daniel Reitz, Tim Langen, and Herwig Ott.
\newblock High-resolution scanning electron microscopy of an ultracold quantum
  gas.
\newblock {\em Nature Physics}, 4(12):949--953, December 2008.

\bibitem{Bakr2009}
Waseem~S. Bakr, Jonathon~I. Gillen, Amy Peng, Simon Foelling, and Markus
  Greiner.
\newblock A quantum gas microscope for detecting single atoms in a
  hubbard-regime optical lattice.
\newblock {\em Nature}, 462(7269):74--U80, November 2009.

\bibitem{Sherson2010}
Jacob~F. Sherson, Christof Weitenberg, Manuel Endres, Marc Cheneau, Immanuel
  Bloch, and Stefan Kuhr.
\newblock Single-atom-resolved fluorescence imaging of an atomic mott
  insulator.
\newblock {\em Nature}, 467(7311):68--U97, September 2010.

\bibitem{Parsons2016}
Maxwell~F. Parsons, Anton Mazurenko, Christie~S. Chiu, Geoffrey Ji, Daniel
  Greif, and Markus Greiner.
\newblock Site-resolved measurement of the spin-correlation function in the
  fermi-hubbard model.
\newblock {\em Science}, 353(6305):1253--1256, 2016.

\bibitem{Boll2016}
Martin Boll, Timon~A. Hilker, Guillaume Salomon, Ahmed Omran, Jacopo Nespolo,
  Lode Pollet, Immanuel Bloch, and Christian Gross.
\newblock Spin- and density-resolved microscopy of antiferromagnetic
  correlations in fermi-hubbard chains.
\newblock {\em Science}, 353(6305):1257--1260, 2016.

\bibitem{Cheuk2016}
Lawrence~W. Cheuk, Matthew~A. Nichols, Katherine~R. Lawrence, Melih Okan, Hao
  Zhang, Ehsan Khatami, Nandini Trivedi, Thereza Paiva, Marcos Rigol, and
  Martin~W. Zwierlein.
\newblock Observation of spatial charge and spin correlations in the 2d
  fermi-hubbard model.
\newblock {\em Science}, 353(6305):1260--1264, 2016.

\bibitem{Read1999}
N.~Read and E.~Rezayi.
\newblock Beyond paired quantum hall states: Parafermions and incompressible
  states in the first excited landau level.
\newblock {\em Physical Review B}, 59(12):8084--8092, 1999.

\bibitem{Kitaev2006a}
A.~Kitaev.
\newblock Anyons in an exactly solved model and beyond.
\newblock {\em Annals of Physics}, 321(1):2--111, January 2006.

\bibitem{Kitaev2003}
A.~Y. Kitaev.
\newblock Fault-tolerant quantum computation by anyons.
\newblock {\em Annals of Physics}, 303(1):PII S0003--4916(02)00018--0, 2003.

\bibitem{Nayak2008}
Chetan Nayak, Steven~H. Simon, Ady Stern, Michael Freedman, and Sankar
  Das~Sarma.
\newblock Non-abelian anyons and topological quantum computation.
\newblock {\em Reviews of Modern Physics}, 80(3):1083--1159, 2008.

\bibitem{Buchler2005}
H.~P. B\"uchler, M.~Hermele, S.~D. Huber, Matthew P.~A. Fisher, and P.~Zoller.
\newblock Atomic quantum simulator for lattice gauge theories and ring exchange
  models.
\newblock {\em Phys. Rev. Lett.}, 95:040402, Jul 2005.

\bibitem{Tagliacozzo2012}
L.~Tagliacozzo, A.~Celi, A.~Zamora, and M.~Lewenstein.
\newblock Optical abelian lattice gauge theories.
\newblock {\em Annals of Physics}, 330:160--191, March 2013.

\bibitem{Wiese2013}
U.-J. Wiese.
\newblock Ultracold quantum gases and lattice systems: quantum simulation of
  lattice gauge theories.
\newblock {\em ANNALEN DER PHYSIK}, 525(10-11):777--796, July 2013.

\bibitem{Zohar2016}
Erez Zohar, J~Ignacio Cirac, and Benni Reznik.
\newblock Quantum simulations of lattice gauge theories using ultracold atoms
  in optical lattices.
\newblock {\em Reports on Progress in Physics}, 79(1):014401--, 2016.

\bibitem{Senthil2004}
T.~Senthil, Ashvin Vishwanath, Leon Balents, Subir Sachdev, and Matthew P.~A.
  Fisher.
\newblock Deconfined quantum critical points.
\newblock {\em Science}, 303(5663):1490--1494, 2004.

\bibitem{Sandvik2007}
Anders~W. Sandvik.
\newblock Evidence for deconfined quantum criticality in a two-dimensional
  heisenberg model with four-spin interactions.
\newblock {\em Phys. Rev. Lett.}, 98:227202, Jun 2007.

\bibitem{Dirac1929}
P.~A.~M. Dirac.
\newblock Quantum mechanics of many-electron systems.
\newblock {\em Proc R Soc Lond A Math Phys Sci}, 123(792):714, April 1929.

\bibitem{Roger1983}
M.~Roger, J.~H. Hetherington, and J.~M. Delrieu.
\newblock Magnetism in solid $^{3}\mathrm{He}$.
\newblock {\em Rev. Mod. Phys.}, 55:1--64, Jan 1983.

\bibitem{Roger1989}
M.~Roger and J.~M. Delrieu.
\newblock Cyclic four-spin exchange on a two-dimensional square lattice:
  Possible applications in high-tc superconductors.
\newblock {\em Phys. Rev. B}, 39:2299--2303, Feb 1989.

\bibitem{Roger2005}
Michel Roger.
\newblock Ring exchange and correlated fermions.
\newblock {\em Journal of Physics and Chemistry of Solids}, 66(8):1412--1416,
  2005.

\bibitem{Weimer2010}
Hendrik Weimer, Markus Muller, Igor Lesanovsky, Peter Zoller, and Hans~Peter
  B\"uchler.
\newblock A rydberg quantum simulator.
\newblock {\em Nat Phys}, 6(5):382--388, May 2010.

\bibitem{Glaetzle2014}
A.~W. Glaetzle, M.~Dalmonte, R.~Nath, I.~Rousochatzakis, R.~Moessner, and
  P.~Zoller.
\newblock Quantum spin-ice and dimer models with rydberg atoms.
\newblock {\em Phys. Rev. X}, 4:041037, Nov 2014.

\bibitem{Surace2019}
Federica~M. Surace, Paolo~P. Mazza, Giuliano Giudici, Alessio Lerose, Andrea
  Gambassi, and Marcello Dalmonte.
\newblock Lattice gauge theories and string dynamics in rydberg atom quantum
  simulators.
\newblock {\em arXiv:1902.09551}, 2019.

\bibitem{Barredo2016}
Daniel Barredo, Sylvain de~Leseleuc, Vincent Lienhard, Thierry Lahaye, and
  Antoine Browaeys.
\newblock An atom-by-atom assembler of defect-free arbitrary two-dimensional
  atomic arrays.
\newblock {\em Science}, 354(6315):1021--, November 2016.

\bibitem{Endres2016}
Manuel Endres, Hannes Bernien, Alexander Keesling, Harry Levine, Eric~R.
  Anschuetz, Alexandre Krajenbrink, Crystal Senko, Vladan Vuletic, Markus
  Greiner, and Mikhail~D. Lukin.
\newblock Atom-by-atom assembly of defect-free one-dimensional cold atom
  arrays.
\newblock {\em Science}, 354(6315):1024--1027, 2016.

\bibitem{Kim2016}
Hyosub Kim, Woojun Lee, Han-gyeol Lee, Hanlae Jo, Yunheung Song, and Jaewook
  Ahn.
\newblock In situ single-atom array synthesis using dynamic holographic optical
  tweezers.
\newblock {\em Nature Comm.}, 7:13317, 2016.

\bibitem{Emery1987}
V.~J. Emery.
\newblock Theory of high-${\mathrm{t}}_{\mathrm{c}}$ superconductivity in
  oxides.
\newblock {\em Phys. Rev. Lett.}, 58:2794--2797, Jun 1987.

\bibitem{Auerbach1998}
Assa Auerbach.
\newblock {\em Interacting Electrons and Quantum Magnetism}.
\newblock Springer, Berlin, 1998.

\bibitem{Toader2005}
A.~M. Toader, J.~P. Goff, M.~Roger, N.~Shannon, J.~R. Stewart, and M.~Enderle.
\newblock Spin correlations in the paramagnetic phase and ring exchange in
  ${\mathrm{la}}_{2}{\mathrm{cuo}}_{4}$.
\newblock {\em Phys. Rev. Lett.}, 94:197202, May 2005.

\bibitem{Paredes2008}
Bel\'en Paredes and Immanuel Bloch.
\newblock Minimum instances of topological matter in an optical plaquette.
\newblock {\em Phys. Rev. A}, 77:023603, Feb 2008.

\bibitem{Dai2017}
Han-Ning Dai, Bing Yang, Andreas Reingruber, Hui Sun, Xiao-Fan Xu, Yu-Ao Chen,
  Zhen-Sheng Yuan, and Jian-Wei Pan.
\newblock Four-body ring-exchange interactions and anyonic statistics within a
  minimal toric-code hamiltonian.
\newblock {\em Nature Physics}, 13:1195--, August 2017.

\bibitem{Haldane1983b}
F.~D.~M. Haldane.
\newblock Continuum dynamics of the 1-d heisenberg anti-ferromagnet -
  identification with the o(3) non-linear sigma-model.
\newblock {\em Physics Letters A}, 93(9):464--468, 1983.

\bibitem{Kennedy1990}
T~Kennedy.
\newblock Exact diagonalisations of open spin-1 chains.
\newblock {\em Journal of Physics: Condensed Matter}, 2(26):5737, 1990.

\bibitem{Kennedy1992}
Tom Kennedy and Hal Tasaki.
\newblock Hidden symmetry breaking and the haldane phase in s=1 quantum spin
  chains.
\newblock {\em Commun. Math. Phys.}, 147:431--484, 1992.

\bibitem{Murphy2009}
M.~Murphy, L.~Jiang, N.~Khaneja, and T.~Calarco.
\newblock High-fidelity fast quantum transport with imperfect controls.
\newblock {\em Phys. Rev. A}, 79:020301(R), 2009.

\bibitem{Muller2009}
M.~M\"uller, I.~Lesanovsky, H.~Weimer, H.~P. B\"uchler, and P.~Zoller.
\newblock Mesoscopic {R}ydberg gate based on electromagnetically induced
  transparency.
\newblock {\em Phys. Rev. Lett.}, 102:170502, 2009.

\bibitem{Fleischhauer2005}
M.~Fleischhauer, A.~Imamoglu, and J.~P. Marangos.
\newblock Electromagnetically induced transparency: Optics in coherent media.
\newblock {\em Reviews of Modern Physics}, 77(2):633--673, 2005.

\bibitem{Haldane1983a}
F.~D.~M. Haldane.
\newblock Non-linear field-theory of large-spin heisenberg anti-ferromagnets -
  semi-classically quantized solitons of the one-dimensional easy-axis neel
  state.
\newblock {\em Physical Review Letters}, 50(15):1153--1156, 1983.

\bibitem{Laeuchli2003}
A.~L\"auchli, G.~Schmid, and M.~Troyer.
\newblock Phase diagram of a spin ladder with cyclic four-spin exchange.
\newblock {\em Phys. Rev. B}, 67:100409, Mar 2003.

\bibitem{Weitenberg2011}
Christof Weitenberg, Manuel Endres, Jacob~F. Sherson, Marc Cheneau, Peter
  Schauss, Takeshi Fukuhara, Immanuel Bloch, and Stefan Kuhr.
\newblock Single-spin addressing in an atomic mott insulator.
\newblock {\em Nature}, 471(7338):319--+, March 2011.

\bibitem{Metavitsiadis2017}
Alexandros Metavitsiadis and Sebastian Eggert.
\newblock Competing phases in spin ladders with ring exchange and frustration.
\newblock {\em Phys. Rev. B}, 95:144415, Apr 2017.

\bibitem{Affleck1987}
Ian Affleck, Tom Kennedy, Elliott~H. Lieb, and Hal Tasaki.
\newblock Rigorous results on valence-bond ground states in antiferromagnets.
\newblock {\em Phys. Rev. Lett.}, 59:799--802, Aug 1987.

\bibitem{Pollmann2012}
Frank Pollmann, Erez Berg, Ari~M. Turner, and Masaki Oshikawa.
\newblock Symmetry protection of topological phases in one-dimensional quantum
  spin systems.
\newblock {\em Phys. Rev. B}, 85:075125, Feb 2012.

\end{thebibliography}
\bibliographystyle{unsrt}

\clearpage
\newpage
\section*{Supplementary information}

\subsection{State preparation}
In the following we present details about the preparation protocol and relevant
experimental parameters that ultimately set the limitations on timescales of
the experiment.

Applying recently developed techniques of rearranging atoms to arbitrary
patterns \cite{Barredo2016, Endres2016, Kim2016}  allows for the
initialization of the desired plaquette pattern with high fidelity.
Raman sideband cooling can be applied to prepare the atoms in the motional
ground state of the tweezers. To achieve a large fraction of ground
state occupation in all three directions, axial confinement of the atoms within
the tweezers is necessary. This can be achieved using a blue-detuned,
higher-order Gaussian beam propagating along the side of the tweezer array.
Strong axial confinement also reduces the position uncertainty of the atoms,
thereby suppressing fluctuations in interactions between the Rydberg states
from run to run.

For the proposed experiments, we suggest encoding the spin and pseudospin
degrees of freedom in the atomic states of \Cs:
\begin{equation}
    \begin{aligned}
        \ket{\tau=0,\,\sigma=\uparrow}&=\ket{F=3,\,m_F=3}\\
        \ket{0, \,\downarrow}&=\ket{3, \,2}\\
        \ket{1, \,\uparrow}&=\ket{4, \,4}\\
        \ket{1, \,\downarrow}&=\ket{4, \,3}
\end{aligned}
\end{equation}

Optical pumping allows for preparation of the full population in a single
hyperfine level such as $\ket{1, \,\uparrow}$, while strong magnetic field
gradients or local optical addressing beams can be used to imprint
state-dependent energy shifts and transfer the spin state on individual sites
using global addressing with microwaves or Raman beams.

\subsection{Trap parameters}
We propose the use of static traps at a far-detuned wavelength to mitigate
heating from off-resonant scattering. $1064\,$nm represents a convenient and
readily available wavelength for these traps. In addition, a different set of
traps at a near-magic wavelength between the $D_1$ and $D_2$ lines of \Cs is
used to transport the $F=4$ hyperfine sublevels, but not $F=3$.
    
An optical trap with $\sigma^+$-polarization and a wavelength of $871.6\,$nm
can strongly shift the $\ket{F=4}$ sublevels but not the $\ket{F=3}$ ones.

\subsubsection{Trap depths}
We assume a static trap depth of $1\,$mK, where good results on ground state
cooling of \Cs is achievable. Assuming a wavelength of $1064\,$nm and a waist
of $w_{\rm stat}\sim 1\,\mu$m, an optical power of around $6\,$mW per trap is
required.

\begin{figure}
    \includegraphics{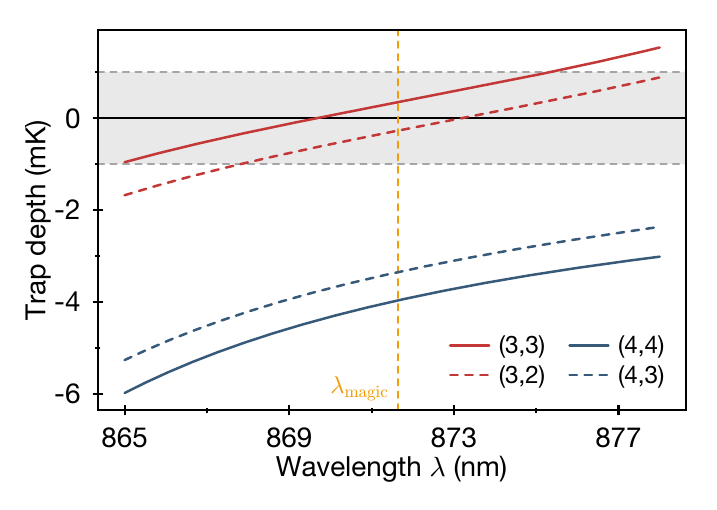}
    \caption{AC Stark shifts on sublevels of \Cs vs. wavelength for the
        rotating traps of $1\,\mu$m waist and $3\,$mW power. Energy shifts on
        $\ket{3,3}$ and $\ket{3,2}$ are smaller than the trap depth of the
        static tweezers (grey-shaded area). The orange dashed line marks the
        optimal wavelength $\lambda_{\rm magic}$ of the rotating traps.}
    \label{fig:RotatingTrapDepth}
\end{figure}

While the static trap depth is flexible, it poses stringent bounds on the depth
of the rotating traps. On the one hand, these should be significantly deeper
than the static traps to allow for efficient relocation of the target qubits in
$\ket{F=4}$. On the other hand, residual light shifts on the $\ket{F=3}$ states
should be small enough as to prevent the atoms in these states from being
removed from the static traps~(Fig.~\ref{fig:RotatingTrapDepth}).

\subsubsection{Scattering rate}
For the trap parameters assumed earlier, the resulting off-resonant photon
scattering rate of the static traps is on the order of $10\,$Hz, which is
much slower than the desired experimental cycle time.

\begin{figure}
    \includegraphics{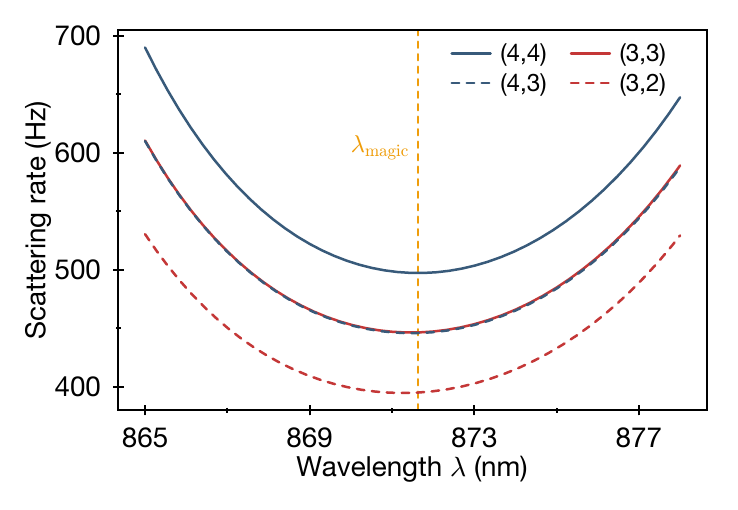}
    \caption{Off-resonant scattering rates of the rotating traps on the
        sublevels of \Cs for the same parameters as
        Fig.~\ref{fig:RotatingTrapDepth}.}
    \label{fig:RotatingScatteringRate}
\end{figure}

For the rotating traps, the near-magic wavelength yields a minimal off-resonant
scattering rate on the order of $500\,$Hz. Therefore it is essential to perform
the rotations sufficiently quickly to avoid destruction of the coherent
superposition of states, heating or scattering into different spin states
altogether. The dependence of scattering rate on the rotating trap wavelength
and the atomic spin state is shown in Fig.~\ref{fig:RotatingScatteringRate}.

\subsubsection{Rotation timescales}
For trap frequencies on the order of $100\,$kHz, a fully adiabatic transport
process would require a timescale much longer than the inverse trap frequency.
The necessary number of operations to implement cyclic ring exchange would need
a very long total trap duration, at which the off-resonant scattering rates
estimated above would severely limit the process fidelity.

Therefore, we propose the implementation of optimized trajectories that allow
for counterdiabatic transport of atoms with no net heating on a timescale of
the inverse energy gap of each trap~\cite{Murphy2009}. Taking into account the
ramping times of the traps, the total transfer period of a single rotation can
take $100\,\mu$s, we can perform on average $\sim20$ rotations until a single
photon is scattered from the traps.

\subsection{Chiral cyclic ring-exchange on a $2\times2$ plaquette}
It is well known that the four-spin cyclic ring-exchange interaction $(\hat{P}_p + \hat{P}_p^\dagger)$ can be written in terms of the spin operators $\hat{\vec{S}}_j$ on the four sites $j=1...4$ of the plaquette (labeled in anti-clockwise direction). To relate our chiral model in Eq.~\eqref{eqCCRdef} to spin models of quantum magnetism, and for the our numerical implementation of the model by DMRG, we also write it out in terms of spin operators. 

As a starting point, we express $\hat{P}_p$ in terms of pairwise spin-permutation operators,
\begin{equation}
\hat{\mathcal{P}}_{ij} = \hat{\mathcal{P}}_{ji} = 2 \l \hat{\vec{S}}_i \cdot \hat{\vec{S}}_j + \frac{1}{4} \r,
\label{eqPijDef}
\end{equation}
for which $\hat{\mathcal{P}}_{ij}  \ket{\sigma_i \sigma_j} = \ket{\sigma_j \sigma_i}$. Hence 
\begin{equation}
\hat{P}_p = \hat{\mathcal{P}}_{43} \hat{\mathcal{P}}_{32} \hat{\mathcal{P}}_{21}.
\end{equation}
Using standard identities for spin operators, we obtain
\begin{multline}
e^{i \phi} \hat{P}_p + e^{- i \phi} \hat{P}_p^\dagger = \cos(\phi) \biggl[ \frac{1}{4} + \hat{\vec{S}}_1 \cdot \hat{\vec{S}}_2 + \hat{\vec{S}}_2 \cdot \hat{\vec{S}}_3  \\
+ \hat{\vec{S}}_3 \cdot \hat{\vec{S}}_4 + \hat{\vec{S}}_4 \cdot \hat{\vec{S}}_1 + \hat{\vec{S}}_2 \cdot \hat{\vec{S}}_4 + \hat{\vec{S}}_1 \cdot \hat{\vec{S}}_3 \\
+ 4 (\hat{\vec{S}}_1 \cdot \hat{\vec{S}}_2 ) (\hat{\vec{S}}_3 \cdot \hat{\vec{S}}_4 )  + 4 (\hat{\vec{S}}_1 \cdot \hat{\vec{S}}_4 ) (\hat{\vec{S}}_2 \cdot \hat{\vec{S}}_3 )  \\
- 4 (\hat{\vec{S}}_1 \cdot \hat{\vec{S}}_3 ) (\hat{\vec{S}}_2 \cdot \hat{\vec{S}}_4 )    \biggr]  + 2 \sin(\phi) \biggl[ \hat{\vec{S}}_1 \cdot (\hat{\vec{S}}_2 \times \hat{\vec{S}}_3) \\
+ \hat{\vec{S}}_1 \cdot (\hat{\vec{S}}_3 \times \hat{\vec{S}}_4) + \hat{\vec{S}}_1 \cdot (\hat{\vec{S}}_2 \times \hat{\vec{S}}_4) + \hat{\vec{S}}_2 \cdot (\hat{\vec{S}}_3 \times \hat{\vec{S}}_4) \biggr].
\end{multline}

\subsection{The spin-$1$ Haldane phase}
To obtain a better understanding of the Haldane phase observed in our DMRG simulations, we perform a rigorous analytical analysis of a simplified model with CCR couplings of strength $K$, $K'$ on alternating plaquettes,
\begin{multline}
\H_{\rm CCR}(K') = K \sum_{p \in \mathsf{P}} ( e^{i \phi} \P_p + e^{ - i \phi}  \P^\dagger_p) \\
+ K' \sum_{p \in \overline{\mathsf{P}}} ( e^{i \phi} \P_p + e^{ - i \phi}  \P^\dagger_p),
\label{eqKKp}
\end{multline}
where $\mathsf{P}$ denotes the set including every second plaquette and $\overline{\mathsf{P}}$ its complement. 

The $K-K'$ model \eqref{eqKKp} can be solved exactly in the limit $K'/K = 0$, where its ground state is a product of decoupled plaquettes $\mathsf{P}$. The eigenstates of a single plaquette $p\in \mathsf{P}$ can be labeled by the total spin $\hat{\vec{S}}_p = \sum_{j=1}^4 \hat{\vec{S}}_j$ where $j=1...4$ denotes the four sites of the plaquette (labeled in anti-clockwise direction).

The two states with $S^z_p = \pm 2$ have an energy $\epsilon_2(\phi) = 2 K \cos \phi$. In the sector with $S^z_p = \pm 1$ there exist four states corresponding to the four positions $j$ of the minority spin. The eigenstates are plane wave superpositions of different $j=1...4$ with discrete momenta $p_n = n \pi/2$ for $n=0,1,2,3$ and corresponding energy $\epsilon_1^n(\phi) = 2 K \cos (\phi + p_n)$. Finally there exist six states with $S^z_p=0$. Four of them correspond to plane-wave superpositions of domain wall configurations, including $\uparrow \uparrow \downarrow \downarrow$ and all cyclic permutations. They have discrete momenta $p_n = n \pi/2$ for $n=0,1,2,3$ and the same energy $\epsilon_2^n(\phi) = 2 K \cos (\phi + p_n)$ as states in the sector $S^z_p=\pm 1$. Two additional states $\ket{\pm}$ correspond to symmetric and anti-symmetric superpositions of N\'eel states $\ket{\uparrow \downarrow \uparrow \downarrow} \pm \ket{ \downarrow \uparrow  \downarrow \uparrow}$ on the plaquette, with eigenenergies  $\epsilon_2^{\pm}(\phi) = 2 K \cos (\phi + q_\pm)$ where $q_+=0$ and $q_-=\pi$.

In the following we focus on the case when $\phi=\pi/2$. The ground state of every plaquette is three-fold degenerate with energy $\epsilon(\pi/2) = -2 K$, and the states are
\begin{flalign}
\ket{\! \uparrow} &= \frac{1}{2} \sum_{j=1}^4 e^{- i j \pi / 2} \hat{S}^-_j \ket{\uparrow \uparrow \uparrow \uparrow} \\
\ket{0} &= \frac{1}{2} \sum_{j=1}^4 e^{- i j \pi / 2} \hat{S}^+_j \hat{S}^+_{j-1} \ket{\downarrow \downarrow \downarrow \downarrow } \\
\ket{\! \downarrow} &= \frac{1}{2} \sum_{j=1}^4 e^{- i j \pi / 2} \hat{S}^+_j \ket{\downarrow \downarrow \downarrow \downarrow}
\end{flalign}
Since the single plaquette is SU$(2)$ invariant, this triplet of states corresponds to the sector $S_p=1$ where the total spin on the plaquette is $\hat{\vec{S}}_p = S_p (S_p + 1)$.

The three states $\ket{\! \uparrow}_p$, $\ket{0}_p$, $\ket{\! \downarrow}_p$ define a system of spin-$1$ operators $\hat{\vec{J}}_p$ on every plaquette $p \in \mathsf{P}$. When $|K'| \ll K$, and without loss of generality $K > 0$, they are protected by a gap $\Delta \approx K$ from further state and $K'$ only introduces coupling between neighboring plaquettes $\langle p,q \rangle$. Making use of SU$(2)$ invariance, we calculated the resulting matrix elements of the term $K' \sum_{p \in \overline{\mathsf{P}}} ( e^{i \phi} \P_p + e^{ - i \phi}  \P^\dagger_p)$ analytically. This leads to the following effective Hamiltonian,
\begin{multline}
\H_{\rm eff} = \sum_{\langle p,q \rangle} \biggl(  \epsilon_0 + \lambda \bigl[ \cos(\theta) \l \hat{\vec{J}}_p \cdot \hat{\vec{J}}_q \r \\
+  \sin(\theta)  \l \hat{\vec{J}}_p \cdot \hat{\vec{J}}_q \r^2 \bigr]  \biggr)
\label{eqHeffHaldane}
\end{multline}
where $\epsilon_0 = - 2 K + \frac{31}{72} K'$. The remaining two coupling constants are given by
\begin{equation}
\lambda \cos(\theta) = \frac{K'}{16}, \qquad \lambda \sin(\theta) = \frac{K'}{144},
\end{equation}
i.e. $\lambda = 0.0629 K'$ and $\theta = 0.035 \pi$.

For these parameters, the effective Hamiltonian is very close to a Heisenberg spin-$1$ chain. Because of the small second term $\propto \lambda \sin(\theta) \ll \lambda \cos(\theta)$, the model interpolates between the exactly solvable AKLT model \cite{Affleck1987} and the simple Heisenberg spin-$1$ chain. Hence the ground state of the effective Hamiltonian \eqref{eqHeffHaldane} is gapped \cite{Haldane1983b} and has spin-$1/2$ edge states \cite{Kennedy1990} reflecting the symmetry protected topological order \cite{Kennedy1992,Pollmann2012}.

We checked numerically by exact diagonalization of numerically accessible system sizes that the system remains gapped at $\phi=\pi/2$ when the ratio $K'/K$ is continuously tuned from $0$ to $1$. Since the system remains inversion symmetric around the central bond of the ladder, and this symmetry is sufficient to protect the topological character of the topological Haldane phase \cite{Pollmann2012}, this establishes that the homogeneous ladder with $K=K'$ is in a non-trivial symmetry-protected phase at $\phi=\pi/2$. The ${\rm SU}(2)$ symmetry of the system is also sufficient to protect the topological Haldane phase \cite{Pollmann2012}.

\subsection{The $J-Q$ model}
Our proposed protocol is versatile enough to implement larger classes of models with multi-spin interactions. In our derivation of Eq.~\eqref{eqHeffCCRderivation} above we only used that fact that $\hat{P}^\dagger \hat{P} = 1$ and the cyclic ring-exchange operator $\hat{P}$ can be replaced by an arbitrary permutation $\mathcal{P}$ of spins. Moreover, introducing more than one control qubit per plaquette allows to implement multiple such terms $\mathcal{P}^{(n)}_p$ per plaquette $p$: For every control atom $n$ associated with plaquette $p$ a coupling term 
\begin{equation*}
\propto \Omega_c^{(n)} \left[ \ket{-,n}_c\bra{+,n} \l 1 + e^{ - i \varphi_c^{(n)}} \hat{\mathcal{P}}^{(n)}_p \r + \hc \right]
\end{equation*}
can be implemented. By integrating out the $n$-th control atom, with detuning $\Delta_c^{(n)} \gg \Omega_c^{(n)}$, an effective Hamiltonian of the form
\begin{equation}
\H_{\rm eff} \propto \sum_{p,n} \frac{(\Omega_c^{(n)})^2}{\Delta_c^{(n)}} \l   e^{- i \varphi_c} \hat{\mathcal{P}}^{(n)}_p  + \hc \r
\end{equation}
is obtained. We envision that control atoms can be stored in a register and moved into the center of the plaquette individually when they are needed for the protocol. 

As a specific example, we discuss an implementation of the $J-Q$ model \cite{Sandvik2007} on a ladder.
The conventional way to write the $J-Q$ Hamiltonian \cite{Sandvik2007} is in terms of spin operators $\hat{\vec{S}}_{\vec{j}}$ on the sites $\vec{j}$ of the lattice,
\begin{equation}
\H_{JQ} = J \sum_{\ij} \hat{\vec{S}}_{\vec{i}} \cdot \hat{\vec{S}}_{\vec{j}} - Q \sum_{\langle \vec{i} \vec{j} \vec{k} \vec{l} \rangle}  \hat{\mathbb{P}}_{\vec{i}, \vec{j}} \hat{\mathbb{P}}_{\vec{k}, \vec{l}},
\label{eqHJQsupp}
\end{equation}
where $\langle \vec{i} \vec{j} \vec{k} \vec{l} \rangle$ denotes a sequence of corners of a plaquette. The second term describes projectors on singlets,
\begin{equation}
\hat{\mathbb{P}}_{\vec{i}, \vec{j}} =  \hat{\vec{S}}_{\vec{i}} \cdot \hat{\vec{S}}_{\vec{j}} - \frac{1}{4}.
\end{equation}
We use a representation in terms of pairwise permutation operators $\hat{\mathcal{P}}_{\vec{i}, \vec{j}} = 2 \l \hat{\vec{S}}_{\vec{i}} \cdot \hat{\vec{S}}_{\vec{j}} + \frac{1}{4} \r$, for which
\begin{multline}
\H_{JQ} =  \frac{2 Q + J}{2}  \sum_{\ij \in R} \hat{\mathcal{P}}_{\vec{i}, \vec{j}}  + \frac{Q+J}{2}  \sum_{\ij \in L} \hat{\mathcal{P}}_{\vec{i}, \vec{j}} \\
- \frac{Q}{4} \sum_{\langle \vec{i} \vec{j} \vec{k} \vec{l} \rangle}  \hat{\mathcal{P}}_{\vec{i}, \vec{j}}  \hat{\mathcal{P}}_{\vec{k}, \vec{l}}
\label{eqDefHJQ}
\end{multline}
up to an overall energy shift, which depends on the boundary conditions. Here $\sum_{\ij \in R}$ ($\sum_{\ij \in L}$) denotes a sum over all links on the rungs (legs) of the ladder and the last term contains a sum over sequences of corners $\langle \vec{i} \vec{j} \vec{k} \vec{l} \rangle$ of the plaquettes.

To implement Eq.~\eqref{eqDefHJQ}, we propose to use one control atom per link $\ij$, to realize the first and second terms $\propto \hat{\mathcal{P}}_{\vec{i}, \vec{j}}$. In addition, two control atoms per plaquette are required to realize $\hat{\mathcal{P}}_{12} \hat{\mathcal{P}}_{34}$ and $\hat{\mathcal{P}}_{14} \hat{\mathcal{P}}_{23}$ respectively; here the sites of the plaquette are labeled by integers $1,2,3,4$ in anti-clockwise direction around the plaquette. 

Similar extensions can be envisioned for implementing the $J-Q$ model in two dimensions. This model features a phase transition around $J / Q \approx 0.04$ between an antiferromagnet and a valence-bond solid, which has been proposed as a candidate for a deconfined quantum critical point \cite{Sandvik2007}.

\end{document}